\documentclass[prd,superscriptaddress,amsfonts,amssymb,amsmath,showpacs,onecolumn]{revtex4-2}
\usepackage{bm}
\usepackage{amsfonts}
\usepackage{latexsym}
\usepackage{graphicx}
\usepackage{amsmath}
\usepackage{palatino}
\usepackage{mathpazo}
\usepackage{textcomp}
\usepackage{float}
\usepackage{booktabs}
\usepackage{dcolumn}
\usepackage{booktabs}
\usepackage{multirow}
\usepackage{hyperref}
\hypersetup{colorlinks,citecolor=blue}
\usepackage{amsmath}
\usepackage{xcolor}
\usepackage{orcidlink}
\usepackage[caption=false]{subfig}
\usepackage{commath}
\def\jnl@style{\it}
\def\aaref@jnl#1{{\jnl@style#1}}

\def\aaref@jnl#1{{\jnl@style#1}}

\def\aj{\aaref@jnl{AJ}}                   
\def\apj{\aaref@jnl{ApJ}}                 
\def\apjl{\aaref@jnl{ApJ}}                
\def\apjs{\aaref@jnl{ApJS}}               
\def\apss{\aaref@jnl{Ap\&SS}}             
\def\aap{\aaref@jnl{A\&A}}                
\def\aapr{\aaref@jnl{A\&A~Rev.}}          
\def\aaps{\aaref@jnl{A\&AS}}              
\def\mnras{\aaref@jnl{Mon.~Not.~Roy.~Astron.~Soc.}}             
\def\prd{\aaref@jnl{Phys.~Rev.~D}}        
\def\prc{\aaref@jnl{Phys.~Rev.~C}}  
\def\prl{\aaref@jnl{Phys.~Rev.~Lett.}}    
\def\qjras{\aaref@jnl{QJRAS}}             
\def\skytel{\aaref@jnl{S\&T}}             
\def\ssr{\aaref@jnl{Space~Sci.~Rev.}}     
\def\zap{\aaref@jnl{ZAp}}                 
\def\nat{\aaref@jnl{Nature}}              
\def\aplett{\aaref@jnl{Astrophys.~Lett.}} 
\def\apspr{\aaref@jnl{Astrophys.~Space~Phys.~Res.}} 
\def\physrep{\aaref@jnl{Phys.~Rep.}}      
\def\physscr{\aaref@jnl{Phys.~Scr}}       
\def\commat{\aaref@jnl{Comm.~Math.~Phys.}}              
\def\science{\aaref@jnl{Science}}               
\def\cqg{\aaref@jnl{Classical Quant.~Grav.}}            
\def\jpcs{\aaref@jnl{JPCS}}                                     
\def\ijmpd{\aaref@jnl{Int.~J.~Mod.~Phys.~D}}                    
\def\grg{\aaref@jnl{Gen.~Relat.~Gravit.}}               
\def\rpp{\aaref@jnl{Rep.~Prog.~Phys.}}          
\def\npa{\aaref@jnl{Nucl.~Phys.~A}}        
\def\lrr{\aaref@jnl{Living Rev.~Rel.}}                   
\def\jcap{\aaref@jnl{J.~Cosmology Astropart.~Phys.}}    
\def\rmp{\aaref@jnl{Rev.~Mod.~Phys.}}   
\def\epjc{\aaref@jnl{Eur.~Phys.~J.~C}}

\allowdisplaybreaks[1]

\begin{document}

\color{black}       

\title{Divergence-free deceleration and energy conditions in non-minimal $f(R,T)$ gravity}

\author{A. Zhadyranova\orcidlink{0000-0003-1153-3438}} 
\email[Email: ]{a.a.zhadyranova@gmail.com}
\affiliation{General and Theoretical Physics Department, L. N. Gumilyov Eurasian National University, Astana 010008, Kazakhstan.}

\author{M. Koussour\orcidlink{0000-0002-4188-0572}}
\email[Email: ]{pr.mouhssine@gmail.com}
\affiliation{Department of Physics, University of Hassan II Casablanca, Morocco.}

\author{Zh. Kanibekova}
\email[Email: ]{kanibekovazhuldyz@gmail.com}
\affiliation{General and Theoretical Physics Department, L. N. Gumilyov Eurasian National University, Astana 010008, Kazakhstan.}

\author{V. Zhumabekova\orcidlink{0000-0002-7223-5373}}
\email[Email: ]{zh.venera@mail.ru}
\affiliation{Theoretical and Nuclear Physics Department, Al-Farabi Kazakh National University, Almaty, 050040, Kazakhstan.}

\author{U. Ismail
\orcidlink{0009-0003-4881-6244}}
\email[Email: ]{umitismail848@gmail.com}
\affiliation{General and Theoretical Physics Department, L. N. Gumilyov Eurasian National University, Astana 010008, Kazakhstan.}

\author{S. Muminov\orcidlink{0000-0003-2471-4836}}
\email[Email: ]{sokhibjan.muminov@gmail.com}
\affiliation{Mamun University, Bolkhovuz Street 2, Khiva 220900, Uzbekistan.}

\author{J. Rayimbaev\orcidlink{0000-0001-9293-1838}}
\email[Email: ]{javlon@astrin.uz}
\affiliation{New Uzbekistan University, Movarounnahr Street 1, Tashkent 100007, Uzbekistan.}
\affiliation{University of Tashkent for Applied Sciences, Str. Gavhar 1, Tashkent 100149, Uzbekistan.}
\affiliation{Urgench State University, Kh. Alimjan Str. 14, Urgench 221100, Uzbekistan.}


\begin{abstract} 
We investigate the divergence-free parametric form of the deceleration parameter within the simplest non-minimal matter-geometry coupling in $f(R,T)$ gravity, where $R$ is the Ricci scalar and $T$ is the trace of the energy-momentum tensor. Specifically, we consider the linear model $f(R,T) = R + 2\lambda T$, where $\lambda$ governs the interaction between matter and geometry. Using this parametric form, we derive the Hubble parameter as a function of redshift $z$ and incorporate it into the modified Friedmann equations. Constraining the model with OHD and Pantheon data, we obtain precise estimates for $H_0$, the present deceleration parameter $q_0$, and its evolutionary component $q_1$, confirming a smooth transition between cosmic deceleration and acceleration. Further, we analyze the evolution of the energy density $\rho$ and total EoS parameter $\omega$ for different $\lambda$ values, highlighting deviations from $\Lambda$CDM and the role of $\lambda$ in shaping cosmic dynamics. In addition, we examine energy conditions, finding that the NEC and DEC are satisfied throughout evolution, while the SEC is violated at late times, supporting the observed acceleration. Our findings demonstrate that this divergence-free parameterization within $f(R,T)$ gravity offers a viable framework for explaining late-time cosmic acceleration while maintaining key observational and theoretical constraints.\\

\end{abstract}

\maketitle

\tableofcontents

\section{Introduction}
\label{sec1}

One of the most profound challenges in modern theoretical physics is explaining the late-time accelerated expansion of the universe. Observational evidence from Type Ia supernovae (SNe Ia) \cite{Riess, Perlmutter}, cosmic microwave background radiation (CMBR) \cite{C.L., R.R.}, baryon acoustic oscillations (BAO) \cite{D.J., W.J.}, Wilkinson microwave anisotropy probe (WMAP) \cite{D.N.}, and large-scale structure (LSS) measurements \cite{E11} strongly support the conclusion that the universe is undergoing accelerated expansion. The prevailing explanation attributes this phenomenon to dark energy (DE), which is believed to dominate the current cosmic energy budget. Among the various DE models, the $\Lambda$CDM model remains the most widely accepted, characterized by a cosmological constant with an equation of state (EoS) $\omega_{\Lambda} = -1$. Despite its success in explaining numerous cosmological observations, the model faces unresolved theoretical challenges. Notably, the coincidence problem questions why the cosmological constant only begins to dominate at late times \cite{Zlatev/1999}. In contrast, the fine-tuning problem highlights the significant discrepancy between the observed and theoretically predicted values of the cosmological constant \cite{Weinberg/1989}. These issues have motivated the exploration of alternative approaches, including dynamical DE models and modified gravity theories, which aim to provide a more fundamental explanation for cosmic acceleration.

An alternative approach to explaining the present acceleration of the universe is to modify the geometric structure of spacetime by extending the Einstein-Hilbert action of general relativity (GR). One of the most well-studied extensions is $f(R)$ gravity \cite{Buchdahl/1970, Dunsby/2010, Carroll/2004}, which generalizes the gravitational action by incorporating an arbitrary function of the Ricci scalar $R$. Other modifications involve non-minimal couplings between matter and geometry, such as $f(R,T)$ gravity \cite{Harko/2011, Koussour_1/2022, Koussour_2/2022, Myrzakulov/2023, KK1}, $f(R,L_m)$ gravity \cite{Harko/2010, Wang/2012, Goncalves/2023, Myrzakulova/2024, Myrzakulov/2024}, $f(\mathcal{T})$ gravity \cite{RR31,RR32,RR33,RR34}, $f(G)$ gravity \cite{Felice/2009, Bamba/2017, Goheer/2009}, $f(Q)$ gravity \cite{Jimenez/2018, Jimenez/2020, Khyllep/2021, MK1, MK2, MK3, MK4, MK5, MK6, MK7, MK8}, and $f(Q,T)$ gravity \cite{Xu/2019, Xu/2020, K6, K7, Bourakadi, KK2, KK3}. Among these, $f(R,T)$ gravity has gained significant attention in recent years, as it introduces direct interactions between the Ricci scalar $R$ and the trace of the energy-momentum tensor $T$, leading to novel cosmological implications \cite{Myrzakulov/2012, Houndjo/2012, Barrientos/2014, Vacaru/2014}. Extensive research has been conducted on various aspects of $f(R,T)$ gravity, covering topics such as thermodynamics \cite{Sharif/2019}, scalar perturbations \cite{Alvarenga/2013}, and wormhole solutions \cite{Moraes/2017}. In addition, within the framework of the Bianchi type-I universe, Sharif and Zubair \cite{Sharif/2012,Sharif/2014} explored cosmological models with a perfect fluid distribution and a massless scalar field. Da Silva et al. \cite{Silva} recently applied $f(R,T)$ gravity to investigate the properties of rapidly rotating neutron stars. Further, Vinutha et al. \cite{Vinutha} examined the field equations and derived the dynamical equations for anisotropic perfect fluid cosmological models. The existence of the G\"{o}del universe in different functional forms of $f(R,T)$ gravity was also analyzed by Bishi et al. \cite{Bishi}.

On the other hand, the existence of an early decelerated expansion phase of the universe is strongly supported by LSS formation and Big Bang nucleosynthesis. To accommodate both the early decelerated and the late-time accelerated expansion stages, the deceleration parameter $q$ must exhibit a signature flip. Furthermore, a transition from a decelerating stage ($q > 0$) to an accelerating stage ($q < 0$) is essential to consistently explain both cosmic structure formation and current observations of accelerated expansion. In recent years, the model-independent method of the universe has been studied in detail using several phenomenological parameterizations of $q(z)$ \cite{Pacif} such as $q(z) = q_0 + q_1 z$ \cite{Riess/2004}, $q(z) = q_0 + q_1 z(1 + z)^{-1}$ \cite{Santos/2011}, $q(z) = \frac{1}{2} + q_1 (1 + z)^{-2}$ \cite{Nair/2012}, $q(z) = q_0 + q_1 \left[ 1 + \ln(1 + z) \right]^{-1}$ \cite{Xu/2008}, $q(z) = \frac{1}{2} + (q_1 z + q_2)(1 + z)^{-2}$ \cite{Gong/2006}, $q(z) = -1 + \frac{3}{2}(1 + z)^{q_2} (q_1 + (1 + z)^{q_2})^{-1}$ \cite{Campo/2012}, $q(z) = q_0 - q_1 (\frac{(1 + z)^{-\alpha} - 1}{\alpha})$ \cite{Mamon/2018}, $q(z) = q_0 + q_1 (\frac{\ln(\alpha + z)}{1 + z} - \beta)$ \cite{Mamon/2017}. However, many of these parameterizations are only valid for $z \ll 1$, while others diverge as $z \to -1$. More recently, a new parameterization of the EoS parameter $\omega_{DE}$ has been introduced, which remains free of divergences in both the past and the future \cite{Barboza/2008,Akarsu/2015}. Despite this progress, only a limited number of such divergence-free parameterizations exist in the literature. Motivated by these considerations, Mamon and Das \cite{Mamon/2016} suggested a simple two-parameter parametric form of $q(z)$ to constrain the evolution of dark energy, ensuring that $q(z)$ remains well-behaved over the entire range $z \in [-1,\infty]$. They analyzed the expansion history of the universe within this divergence-free framework. The functional form of this $q$-parametrization closely resembles the one introduced in \cite{Barboza/2008} for the EoS parameter $\omega_{DE}$.

In this work, we explore a divergence-free parametric form of the deceleration parameter within the simplest non-minimal matter-geometry coupling in $f(R,T)$ gravity, adopting the functional form $f(R,T) = R + 2\lambda T$, where $\lambda$ dictates the interaction between matter and geometry. Using this parameterization, we derive the Hubble parameter as a function of redshift $z$ and incorporate it into the modified Friedmann equations to analyze the energy conditions. The structure of this paper is as follows: Sec. \ref{sec2} introduces the overview of $f(R,T)$ gravity. In Sec. \ref{sec3}, we present the modified Friedmann equations and derive the Hubble parameter using the divergence-free parametric form. Sec. \ref{sec4} discusses the observational data used to constrain the model parameters and examines the evolution of cosmological parameters, including the dimensionless energy density and total EoS parameter. The analysis of energy conditions, namely NEC, WEC, DEC, and SEC, is conducted in Sec. \ref{sec5}. Finally, Sec. \ref{sec6} provides a summary of our findings.

\section{Overview of $f(R,T)$ Modified Gravity}
\label{sec2}

In this section, we provide a concise overview of the field equations in $f(R,T)$ modified gravity and examine the associated violation of the energy-momentum tensor. The action for $f(R,T)$ gravity was originally introduced as \cite{Harko/2011}
\begin{equation}
\mathbb{S}=\frac{1}{2}\int  f(R,T)\sqrt{-g}d^{4}x +\int L_{m}\sqrt{-g}d^{4}x.\label{action}
\end{equation}

Here, we adopt the conventions $8\pi G = c = 1$. The function $f(R, T)$ represents an arbitrary functional dependence, where $R$ denotes the Ricci scalar, $T$ is the trace of the energy-momentum tensor defined as $T = g_{\mu\nu} T^{\mu\nu}$, and $L_m$ is the matter Lagrangian density. The energy-momentum tensor, which characterizes the distribution and flow of energy and momentum within spacetime, is given by  
\begin{equation}\label{1e}
T_{\mu\nu} = \frac{-2}{\sqrt{-g}} \frac{\delta(\sqrt{-g}L_m)}{\delta g^{\mu\nu}}.
\end{equation}

Furthermore, the Ricci scalar $R$, which measures the overall curvature of spacetime, is obtained by contracting the Ricci tensor $R_{\mu\nu}$. The Ricci tensor itself quantifies the extent to which the volume of a geodesic ball in curved spacetime deviates from its counterpart in flat spacetime. This relationship is expressed as   
\begin{equation}\label{1b}
R= g^{\mu\nu} R_{\mu\nu},
\end{equation} 
where
\begin{equation}\label{1c}
R_{\mu\nu}= \partial_\lambda \Gamma^\lambda_{\mu\nu} - \partial_\mu \Gamma^\lambda_{\lambda\nu} + \Gamma^\lambda_{\mu\nu} \Gamma^\sigma_{\sigma\lambda} - \Gamma^\lambda_{\nu\sigma} \Gamma^\sigma_{\mu\lambda}.
\end{equation}

The components $\Gamma^\alpha_{\beta\gamma}$, referred to as the Levi-Civita connection, characterize the manner in which vectors are parallel transported along curves within curved spacetime. These components are expressed in terms of the metric tensor $g_{\mu\nu}$ and its first derivatives as:  
\begin{equation}\label{4}
\Gamma^\alpha_{\beta\gamma}= \frac{1}{2} g^{\alpha\lambda} \left( \frac{\partial g_{\gamma\lambda}}{\partial x^\beta} + \frac{\partial g_{\lambda\beta}}{\partial x^\gamma} - \frac{\partial g_{\beta\gamma}}{\partial x^\lambda} \right).
\end{equation}

By varying the action (\ref{action}) with respect to the metric tensor $g_{\mu\nu}$, the corresponding field equations, which govern the dynamics of gravitational interactions, are derived \cite{Harko/2011}:
\begin{equation} \label{field}
f_{R}(R,T)R_{\mu\nu} - \frac{1}{2} f(R,T) g_{\mu\nu}+ \left(g_{\mu\nu}\Box - \nabla_{\mu}\nabla_{\nu}\right) f_{R}(R,T) = T_{\mu\nu} - f_{T}(R,T) T_{\mu\nu} - f_{T}(R,T) \Theta_{\mu\nu},
\end{equation}
where 
\begin{equation}
\Theta_{\mu\nu} \equiv g^{\alpha\beta}\frac{\delta T_{\alpha\beta}}{\delta g^{\mu\nu}}=-2T_{\mu\nu} +g_{\mu\nu}L_m -2g^{\alpha\beta}\frac{\partial^2 L_m}{\partial g^{\mu\nu}\partial g^{\alpha \beta}}.
\end{equation}

Also, we adopt the following definitions for simplicity:
\begin{equation}
f \equiv f(R,T), \quad f_R \equiv \frac{\partial f(R,T)}{\partial R}, \quad f_T \equiv \frac{\partial f(R,T)}{\partial T}.
\end{equation}

Moreover, applying the covariant derivative to Eq. (\ref{field}) yields the following relation:
\begin{equation}
\label{NC}
\nabla^{\mu}T_{\mu\nu}=\frac{f_T}{8\pi -f_T}[(T_{\mu\nu}+\Theta_{\mu\nu})\nabla^{\mu}\ln f_T +\nabla^{\mu}\Theta_{\mu\nu}-(1/2)g_{\mu\nu}\nabla^{\mu}T].
\end{equation}

Thus, the equation above exemplifies a fundamental feature of the $f(R,T)$ gravity framework, highlighting a deviation from the standard conservation of the matter-energy-momentum tensor. In general relativity, the covariant divergence of this tensor vanishes, ensuring strict conservation laws. However, in $f(R,T)$ gravity, this property no longer holds, as indicated by $\nabla^\mu T_{\mu \nu} \neq 0$. This non-conservation implies the presence of an additional force acting on matter, leading to deviations from geodesic motion. Physically, it represents energy exchange within a given volume, which may be interpreted as particle creation or dissipation effects. The non-zero divergence of the energy-momentum tensor signifies the influence of the trace-dependent modifications introduced by the theory. Nevertheless, when the function $f(R,T)$ lacks explicit dependence on $T$, the standard conservation law is recovered, restoring the conventional dynamics of matter fields \cite{Harko/2011,KK1}.

\section{Cosmological model and solutions} \label{sec3}

Here, we apply the spatially flat Friedmann-Lema\^itre-Robertson-Walker metric according to the cosmological principle. In line with this principle, our universe is homogeneous and isotropic on a large scale, meaning it looks the same in every direction and at every point \cite{ryden/2003},
\begin{equation}
ds^2=-dt^2+a^2 (t) \left[dr^2+ r^2 (d\theta^2+sin^2\theta d\phi^2 ) \right], \label{FLRW}
\end{equation} 
where $a(t)$ represents the scale factor of the universe, which characterizes the evolution of distances between points over time. Furthermore, using the metric (\ref{FLRW}), the Ricci scalar is given by $R= 6 ( \dot{H}+2H^2)$. Here, $H = \frac{\dot{a}}{a}$ is the Hubble parameter, which quantifies the rate of expansion of the universe.

Furthermore, we assume the matter content of the universe to be a perfect fluid, for which the energy-momentum tensor is given by
\begin{equation}\label{2c}
T_{\mu\nu}=(\rho+p)u_\mu u_\nu + pg_{\mu\nu},
\end{equation}
where $\rho$ is the energy density, $p$ is the pressure, $u_\mu$ is the four-velocity of the fluid, and $g_{\mu\nu}$ is the metric tensor. Therefore, the trace of Eq. (\ref{2c}) is given by $T = 3p-\rho$. Also, we assume the Lagrangian of the perfect fluid to be $L_m =p$. The corresponding term $\Theta_{\mu\nu}$ is expressed as $\Theta_{\mu\nu}= -2 T_{\mu\nu}+ p g_{\mu\nu}$.

By applying the FLRW metric to Eq. (\ref{FLRW}), we obtain the two modified Friedmann equations as \cite{Shabani/2018}
\begin{equation} \label{F1}
3H^2 f_R + \frac{1}{2} \left( f - f_R R \right) + 3 \dot{f_R} H = \left( 1 + f_T \right) \rho + f_T p,
\end{equation}
\begin{equation} \label{F2}
2 f_R \dot{H} + \ddot{f_R} - \dot{f_R} H = - \left( 1 + f_T \right)(\rho+p),
\end{equation}
where dot (.) denotes the time derivative and $f_R$, and $f_T$ are the derivative with respect to $R$ and $T$, respectively.

In this study, we consider the functional form $f(R,T) = R + 2 \lambda T$, where $\lambda$ is a constant \cite{Harko/2011}. The choice is motivated by its simplicity and physical relevance. It ensures second-order field equations, avoiding higher-order complexities. The term $2\lambda T$ introduces a matter-geometry interaction, leading to a non-conserved energy-momentum tensor, which can be interpreted as energy exchange. This form is also phenomenologically viable, offering straightforward applications in cosmology, where it can mimic DE effects. While alternative choices exist, they introduce additional complexities. Thus, this form balances analytical tractability with key modifications to GR. Notably, setting $\lambda = 0$ recovers the standard GR with $f(R,T) = R$. Further, for $T = 0$, the theory reduces to $f(R)$ gravity, which is equivalent to GR and satisfies all solar system observational constraints. With the selected form of $ f(R,T)$ and by solving Eqs. (\ref{F1}) and (\ref{F2}), we derive the expressions for the pressure $p$ and energy density $\rho$ as follows:
\begin{equation} \label{p}
p = -\frac{(3+6\lambda)}{(1+3\lambda)^2 - \lambda^2} H^2 - \frac{2(1+3\lambda)}{(1+3\lambda)^2 - \lambda^2} \dot{H},
\end{equation}
\begin{equation} \label{rho}
\rho = \frac{(3+6\lambda)}{(1+3\lambda)^2 - \lambda^2} H^2 - \frac{2\lambda}{(1+3\lambda)^2 - \lambda^2} \dot{H}.
\end{equation}

However, when $\lambda = 0$, the newly modified Friedmann equations lead to the standard Friedmann equations: $p = -(3 H^2 +2 \dot{H})$ and $\rho = 3 H^2$. For simplicity's sake, we assume
\begin{equation}
A = \frac{(3+6\lambda)}{(1+3\lambda)^2 - \lambda^2}, \quad B=\frac{2(1+3\lambda)}{(1+3\lambda)^2 - \lambda^2}, \quad C=\frac{2\lambda}{(1+3\lambda)^2 - \lambda^2}.
\end{equation}

Therefore, Eqs. (\ref{p})-(\ref{rho}) become
\begin{equation}
p = -A H^2 -B \dot{H}.
\end{equation}
\begin{equation}
\rho = A H^2 - C \dot{H}.
\end{equation}

Moreover, the total EoS parameter $\omega =\frac{p}{\rho}$ can be written as
\begin{equation} \label{w}
\omega =\frac{-A H^2 -B \dot{H}}{A H^2 - C \dot{H}}=\frac{-A -B \frac{\dot{H}}{H^2}}{A - C \frac{\dot{H}}{H^2}}.    
\end{equation}

The deceleration parameter is defined as
\begin{equation} \label{q}
q = -1 - \frac{\dot{H}}{H^2}.    
\end{equation}

Here, $q > 0$ signifies a decelerating phase, while $q < 0$ indicates an accelerating phase of the universe. Integrating Eq. (\ref{q}) gives:
\begin{equation} \label{qH}
H(z) = H_{0}\, exp\Big(\int_{0}^{z} \frac{1+ q(z)}{(1 + z)} dz \Big),
 \end{equation}
where $H_0=H(z=0)$ represents the Hubble parameter at the current epoch, and the redshift $z$, defined as $z = \frac{\lambda_{\text{obs}} - \lambda_{\text{emit}}}{\lambda_{\text{emit}}} = \frac{1}{a(t)} - 1$, measures the relative change in the wavelength of light due to cosmic expansion. If the functional form of $q(z)$ is specified, it provides findings into the evolution of both the Hubble parameter and the EoS parameter. In this context, we focus on a divergence-free parametric form of the deceleration parameter, expressed as \cite{Mamon/2016,Hanafy/2019}
\begin{equation}
\label{qz}
	q(z) = q_{0}+q_{1}\frac{z(1+z)}{1+z^2}.
\end{equation}
where $q_0$ represents the present-day value of $q$, and $q_1$ quantifies its variation with redshift $z$. This parametrization has two key asymptotic behaviors: (i) At high redshifts ($z \gg 1$), $q(z)$ asymptotically approaches $q_0 + q_1$, describing the early universe. (ii) At low redshifts ($z \ll 1 $), $q(z)$ reduces to $q_0 + q_1 z$, capturing the recent evolution of cosmic acceleration. An important feature of this model is that it avoids divergences, yielding finite values of $q$ over the entire redshift range, $z \in [-1, \infty]$, thereby covering the full cosmological timeline, including the future ($z = -1$). Furthermore, this form draws inspiration from a widely adopted divergence-free parametrization of $\omega_{DE}$ \cite{Barboza/2008}, demonstrating its flexibility in reproducing the deceleration parameter's behavior for a broad class of accelerating cosmological models.

Substituting Eq. (\ref{qz}) into Eq. (\ref{qH}), we derive the expression for the Hubble parameter in terms of $z$ as follows: 
\begin{equation} \label{Hz}
H(z) = H_{0} (1+z)^{(1+ q_{0})} (1+z^2)^{\frac{q_1}{2}}.
\end{equation}

From Eq. (\ref{q}), solving for $\frac{\dot{H}}{H^2}$ and substituting the parametric form of $q(z)$, we obtain  
\begin{equation}
\frac{\dot{H}}{H^2} = -(1 + q) = -\left(1 + q_0 + q_1 \frac{z (1+z)}{1+z^2}\right).
\end{equation}

Now, substituting $\frac{\dot{H}}{H^2}$ into Eq. (\ref{w}), we obtain
\begin{equation}
\omega(z) =\frac{-A +B \left(1 + q_0 + q_1 \frac{z (1+z)}{1+z^2}\right)}{A + C \left(1 + q_0 + q_1 \frac{z (1+z)}{1+z^2}\right)}.    
\end{equation}

In the next section, our goal is to validate this model. To this end, we usee various cosmological datasets, including observational Hubble data and SNe Ia from the Pantheon sample, to constrain the parameters $H_0$, $q_0$, and $q_1$.

\section{Observational data}
\label{sec4}

In this section, we evaluate the proposed model against contemporary observational datasets, specifically the observational Hubble data (OHD) \cite{Yu/2018, Moresco/2015} and the Pantheon SNe Ia sample \cite{Scolnic/2018}, which includes 1048 data points from various surveys, such as the Pan-STARRS1 Medium-Deep Survey (PS1-MDS), the Sloan Digital Sky Survey (SDSS), the Supernova Legacy Survey (SNLS), and additional low-redshift and Hubble Space Telescope (HST) samples \cite{Chang/2019}.

\subsection{OHD}
A set of 31 measurements of the Hubble parameter $H(z)$, derived using the differential age method (DA), is used to estimate the universe's expansion rate \cite{Yu/2018, Moresco/2015}. The Hubble parameter is calculated as $H(z) = -\frac{1}{(1+z)}\frac{dz}{dt}$. The parameters are optimized by minimizing the chi-square function,
\begin{equation}
    \chi^2_{OHD} = \sum_{i=1}^{31} \frac{\left[H(z_i, H_0,q_0,q_1) - H_{\text{obs}}(z_i)\right]^2}{\sigma(z_i)^2},
\end{equation}
where $H(z_i, H_0,q_0,q_1)$ is the theoretical value at redshift $z_i$, and $H_{obs}(z_i)$ and $\sigma(z_i)^2$ denote the observed value and its uncertainty, respectively. 

\subsection{Pantheon data}
The Pantheon dataset, the largest spectroscopically confirmed compilation of SNe Ia, consists of 1048 data points of apparent magnitude versus redshift spanning the range $0.01 < z < 2.3$ \cite{Scolnic/2018, Chang/2019}. It incorporates 279 SNe Ia discovered by the PS1-MDS, along with distance measurements from the SDSS, SNLS, and additional low-redshift and HST samples. The luminosity distance $d_L(z)$ in a spatially flat universe is expressed as \cite{Planck/2018},  
\begin{equation}
d_L(z) = c(1+z)\int^z_0 \frac{d\overline{z}}{H(\overline{z})},  
\end{equation}  
where $c$ is the speed of light. The theoretical distance modulus is given by $\mu^{th} = 5\log_{10}d_L(z) + \mu_0$, where $\mu_0 = 5 \log_{10} \frac{1}{H_0 Mpc} + 25$. The chi-square function for the Pantheon data is expressed as,  
\begin{equation}
\chi^2_{Pantheon} = \sum_{i,j=1}^{1048} \Delta\mu_i \left(C^{-1}_{Pantheon}\right)_{ij} \Delta\mu_j,  
\end{equation}  
where $\Delta\mu_i = \mu^{th} - \mu_i^{obs}$, with $\mu^{th}$ representing the theoretical distance modulus and $\mu_i^{obs}$ its observed counterpart. 

\subsection{Joint analysis}
To obtain joint constraints on parameters $H_0$, $q_0$ and $q_1$, the combined likelihood and chi-square functions are defined as
\begin{eqnarray}
\mathcal{L}_{Joint} &=& \mathcal{L}_{OHD} \times \mathcal{L}_{Pantheon},\\
\chi^2_{Joint} &=& \chi^2_{OHD} + \chi^2_{Pantheon}.
\end{eqnarray}

\begin{figure}[H]
\centering
\includegraphics[scale=0.7]{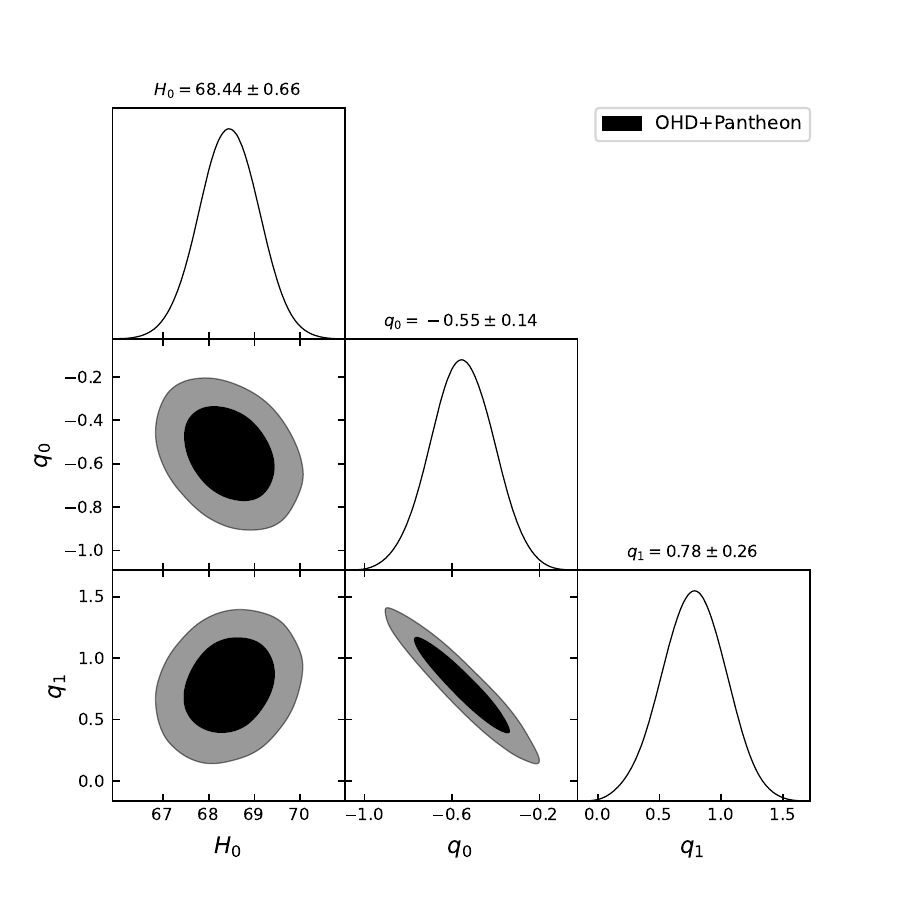}
\caption{This figure depicts the corner plot for the divergence-free parametric form of the deceleration parameter, displaying the one-dimensional marginalized posterior distributions and the two-dimensional confidence contours for the parameters $H_0$, $q_0$, and $q_1$ using the joint datasets.}
\label{F_Com}
\end{figure}

The Markov Chain Monte Carlo (MCMC) method, implemented using the \texttt{emcee} library, is employed to minimize $\chi^2$ \cite{emcee}. Fig. \ref{F_Com} presents a corner plot for the divergence-free parametric form of the deceleration parameter, showing the one-dimensional marginalized posterior distributions and the two-dimensional confidence contours for $H_0$, $q_0$, and $q_1$. The best-fit value for the Hubble parameter is $H_0 = 68.44 \pm 0.66$, indicating a precise determination using the combined OHD and Pantheon data. The current deceleration parameter is $q_0 = -0.55 \pm 0.14$ \cite{Planck/2018,Mamon/2017}, confirming the universe's accelerated expansion ($ q_0 < 0$), while $q_1 = 0.78 \pm 0.26$ describes the redshift evolution of $q(z)$. The confidence contours reveal mild correlations among the parameters: $H_0$ and $q_0$ show a slight negative correlation, $H_0$ and $q_1$ are largely uncorrelated, and $q_0$ and $q_1$ exhibit a moderate negative correlation, linking the current acceleration to its evolutionary rate. These results support a smooth and finite evolution of $q(z)$ across the redshift range, consistent with the accelerating universe scenario and the divergence-free nature of the model.

Based on these values from observational data, we discuss the behavior of the energy density and the EoS parameter for different coupling parameter values $\lambda$ within the $f(R,T)$ gravity framework. The parameter $\lambda$ represents the degree of the interaction between matter and geometry.

\begin{figure}[H]
   \begin{minipage}{0.48\textwidth}
     \centering
     \includegraphics[width=\linewidth]{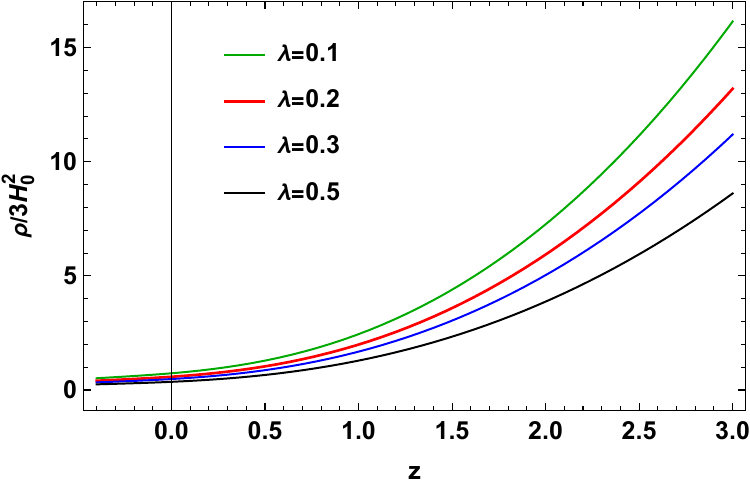}
     \caption{The evolution of the dimensionless energy density, $\rho / (3H_0^2)$ with respect to redshift $z$ for different values of $\lambda$.}\label{F_rho}
   \end{minipage}\hfill
   \begin{minipage}{0.48\textwidth}
     \centering
     \includegraphics[width=\linewidth]{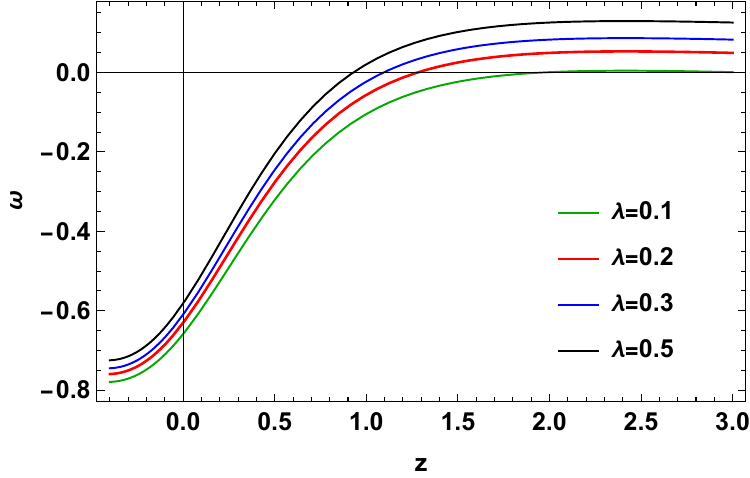}
     \caption{The evolution of the total EoS parameter $\omega$ with respect to redshift $z$ for different values of $\lambda$.}\label{F_w}
   \end{minipage}
\end{figure}

Fig. \ref{F_rho} shows the evolution of the dimensionless energy density, $\rho / (3H_0^2)$, as a function of redshift $z$ for different values of $\lambda$. The curves correspond to $\lambda = 0.1, 0.2, 0.3, 0.5$, represented by green, red, blue, and black, respectively. As the redshift $z$ increases, corresponding to earlier epochs in the universe's history, the energy density grows, reflecting the expected behavior during the matter-dominated era. The figure shows that lower values of $\lambda$ (e.g., $\lambda = 0.1$, green curve) lead to an increased energy density compared to higher values (e.g., $\lambda = 0.5$, black curve) at the same redshift. This behavior indicates that the coupling parameter $\lambda$ amplifies the contribution of matter to the energy density as the redshift increases. The divergence-free parametric form of the deceleration parameter ensures a smooth transition between the decelerating and accelerating phases of the universe's expansion. This interaction within the $f(R,T)$ gravity model modifies the energy density evolution compared to the standard $\Lambda$CDM model.

The EoS parameter plays a crucial role in characterizing the different epochs of the universe's accelerated and decelerated expansion. These epochs include: $\omega = 1$, corresponding to a stiff fluid; $\omega = 1/3$, associated with the radiation-dominated phase; and $\omega = 0$, indicative of the matter-dominated phase. For an accelerating universe ($\omega < -1/3$), the EoS parameter describes one of three possible states: the cosmological constant ($\omega = -1$), quintessence ($-1 < \omega < -1/3$), or the phantom era ($\omega < -1$) \cite{Myrzakulov/2023}. Fig. \ref{F_w} displays the behavior of the total EoS parameter, $\omega$, as a function of redshift $z$ for the linear $f(R,T)$ gravity model with the divergence-free parametric form of the deceleration parameter, where the curves correspond to different values of the coupling parameter $\lambda$ ($\lambda = 0.1, 0.2, 0.3, 0.5$). At higher redshifts ($z > 1.5$), $\omega$ approaches zero, corresponding to the matter-dominated era ($\omega \approx 0$), confirming that the model recovers standard cosmological behavior in the early universe, irrespective of $\lambda$. As $z \to 0$, $\omega$ decreases to negative values, reflecting late-time cosmic acceleration, with smaller $\lambda$ values (e.g., $\lambda = 0.1$) leading to more negative $\omega$, indicating stronger deviations from $\Lambda$CDM and enhanced acceleration effects, while larger $\lambda$ values (e.g., $\lambda = 0.5$) result in $\omega$ values closer to zero, implying weaker matter-geometry coupling. The smooth transition of $\omega$ from $\omega \approx 0$ in the matter-dominated era to $\omega < -\frac{1}{3}$ in the quintessence phase ensures a divergence-free parameterization and aligns well with observational constraints, highlighting how $f(R,T)$ gravity naturally explains cosmic acceleration, with $\lambda$ being the key factor shaping the EoS dynamics.

\section{Energy conditions} \label{sec5}

Energy conditions are crucial in analyzing the geodesic structure of the universe and are derived from the Raychaudhuri equations \cite{Raychaudhuri}. For a congruence of timelike and null geodesics, these equations take the forms \citep{Ehlers/2006,Nojiri/2007},
\begin{equation}\label{12}
\frac{d\theta}{d\tau}=-\frac{1}{3}\theta^2-\sigma_{\mu\nu}\sigma^{\mu\nu}+\omega_{\mu\nu}\omega^{\mu\nu}-R_{\mu\nu}u^{\mu}u^{\nu}\,,
\end{equation}
and 
\begin{equation}\label{13}
\frac{d\theta}{d\tau}=-\frac{1}{2}\theta^2-\sigma_{\mu\nu}\sigma^{\mu\nu}+\omega_{\mu\nu}\omega^{\mu\nu}-R_{\mu\nu}n^{\mu}n^{\nu}\,,
\end{equation}
where $\theta$ denotes the expansion scalar, $n^{\mu}$ is the null vector, and $\sigma_{\mu\nu}$ and $\omega_{\mu\nu}$ represent the shear and rotation tensors associated with the vector field $u^{\mu}$. The conditions for attractive gravity require that  
\begin{equation}\label{14}
R_{\mu\nu}u^{\mu}u^{\nu}\geq0 \, ; \,\,\,\,\,
 R_{\mu\nu}n^{\mu}n^{\nu}\geq0\,.
\end{equation}

In the context of $f(R,T)$ gravity with a perfect fluid matter distribution, these constraints lead to the following energy conditions:
\begin{itemize}
    \item \textbf{Null energy condition (NEC):} $\rho + p \geq 0$
    \begin{equation}
    \rho + p = (A H^2 - C \dot{H}) + (-A H^2 - B \dot{H}) = -(B + C) \dot{H}.
    \end{equation}
    The NEC requires $-(B + C) \dot{H} \geq 0$. Therefore, the condition is satisfied if $\dot{H} \geq 0$ and $B + C \leq 0$.
    
    \item \textbf{Weak energy condition (WEC):} $\rho \geq 0, \ \rho + p \geq 0$
    \begin{itemize}
        \item The first part: $\rho = A H^2 - C \dot{H} \geq 0$, which implies $A H^2 \geq C \dot{H}$.
        \item The second part: $\rho + p = -(B + C) \dot{H} \geq 0$, as shown above, requiring $\dot{H} \geq 0$ and $B + C \leq 0$.
    \end{itemize}
    Therefore, the WEC is satisfied if $A H^2 \geq C \dot{H}$, $\dot{H} \geq 0$, and $B + C \leq 0$.
    
    \item \textbf{Dominant energy condition (DEC):} $\rho \geq 0, \ |p| \leq \rho$
    \begin{itemize}
        \item The condition $\rho \geq 0$ gives $A H^2 \geq C \dot{H}$, as before.
        \item The condition $|p| \leq \rho$ translates into:
        \begin{equation}
        |-A H^2 - B \dot{H}| \leq A H^2 - C \dot{H}.
        \end{equation}
        This inequality needs to be analyzed in terms of the coefficients $A$, $B$, and $C$ along with the values of $H^2$ and $\dot{H}$.
    \end{itemize}
    
    \item \textbf{Strong energy condition (SEC):} $\rho + 3p \geq 0$
    \begin{equation}
    \rho + 3p = (A H^2 - C \dot{H}) + 3(-A H^2 - B \dot{H}) = -2A H^2 - (3B+C) \dot{H}.
    \end{equation}
    For acceleration, we require SEC violation \cite{Visser/2000}, which means: $\rho + 3p < 0 \quad \Rightarrow \quad -2A H^2 - (3B+C) \dot{H}< 0$. This implies that the cosmic acceleration phase occurs when: $-2A H^2 < (3B+C) \dot{H}$. Physically, this means that the contribution from the time variation of the Hubble parameter ($\dot{H}$) must be large in magnitude, compared to $H^2$, to drive an accelerated expansion. This is typical in scenarios involving dark energy, inflation, or modified gravity theories, where the effective EoS satisfies $\omega < -\frac{1}{3}$.
\end{itemize}

\begin{figure}[H]
   \begin{minipage}{0.32\textwidth}
     \centering
     \includegraphics[width=\linewidth]{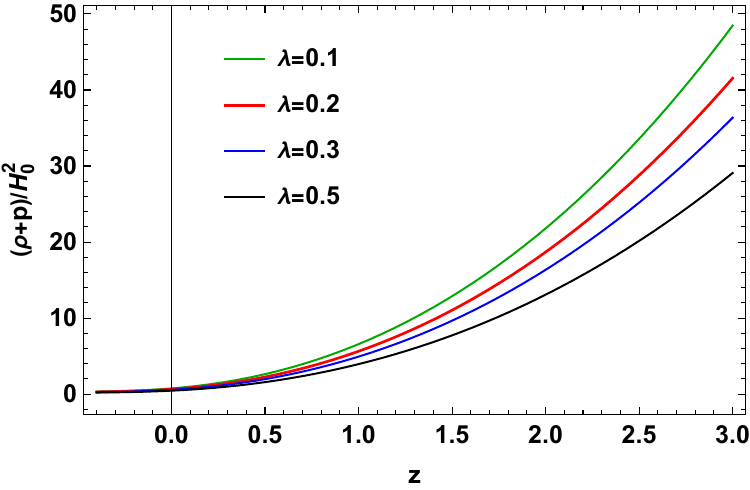}
     \caption{The evolution of NEC condition $(\rho + p)/H_0^2$ with respect to redshift $z$ for different values of $\lambda$.}\label{F_NEC}
   \end{minipage}\hfill
   \begin{minipage}{0.32\textwidth}
     \centering
     \includegraphics[width=\linewidth]{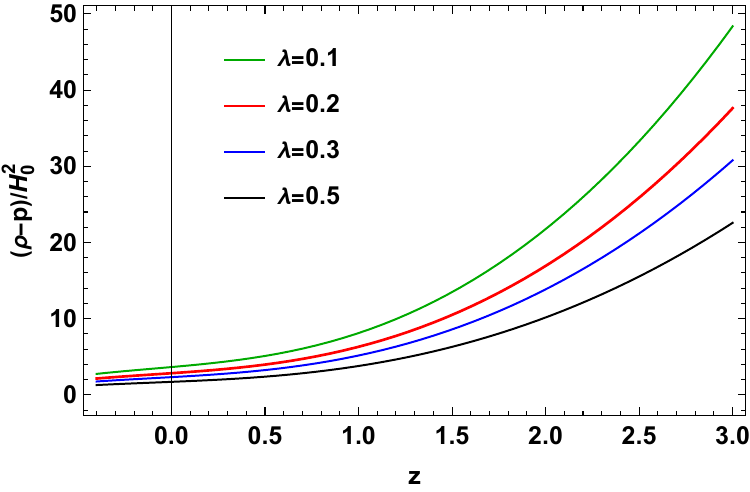}
     \caption{The evolution of DEC condition $(\rho - p)/H_0^2$ with respect to redshift $z$ for different values of $\lambda$.}\label{F_DEC}
   \end{minipage}\hfill
   \begin{minipage}{0.32\textwidth}
     \centering
     \includegraphics[width=\linewidth]{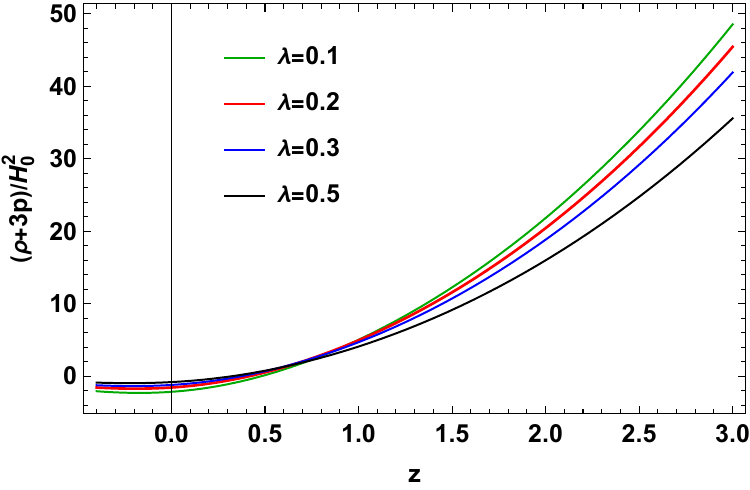}
     \caption{The evolution of SEC condition $(\rho + 3p)/H_0^2$ with respect to redshift $z$ for different values of $\lambda$.}\label{F_SEC}
   \end{minipage}
\end{figure}

Fig. \ref{F_NEC} shows the evolution of $(\rho + p)/H_0^2$ with respect to the redshift $z$ for different values of $\lambda$ in the $f(R,T)$ gravity framework. The NEC, $(\rho + p) \geq 0$, is satisfied for all parameter values, as the curves remain non-negative throughout the redshift range. At low redshifts ($z \sim 0$), the values of $(\rho + p)/H_0^2$ approach zero, indicating that the NEC is satisfied in the present. At higher redshifts ($z > 1$), the values increase significantly to large positive values. Increasing $\lambda$ reduces $(\rho + p)/H_0^2$ at fixed $z$, demonstrating its moderating effect on the effective energy density and pressure. Furthermore, Fig. \ref{F_DEC} shows the behavior of $(\rho - p)/H_0^2$ as a function of the redshift $z$ for different values of $\lambda$ in the $f(R,T)$ gravity framework. The figure confirms that the DEC, $(\rho - p) \geq 0$, is satisfied throughout the entire redshift range for all values of $\lambda$. At low redshifts ($z \sim 0$), $(\rho - p)/H_0^2$ begins with small positive values and increases steadily with $z$. A higher $\lambda$ leads to a reduction in $(\rho - p)/H_0^2$ at fixed $z$, showcasing its impact on moderating the effective energy density and pressure. Finally, Fig. \ref{F_SEC} shows the behavior of $(\rho + 3p)/H_0^2$ as a function of the redshift $z$ for various values of $\lambda$ within the $f(R,T)$ gravity framework. It explores the violation of the SEC, $(\rho + 3p) < 0$. The results reveal that $(\rho + 3p)/H_0^2$ remains negative throughout the considered redshift range ($z \leq 0$), implying that the SEC is violated in both the present and future epochs. At higher redshifts ($z > 0$), $(\rho + 3p)/H_0^2$ increases monotonically to positive values, corresponding to the dominance of matter and radiation in the early universe. As $\lambda$ increases, the curves show a slight suppression, reflecting the impact of the $\lambda T$ term on altering effective gravitational dynamics. At $z \sim 0$ (the present epoch), $(\rho + 3p)/H_0^2$ approaches lower negative values, signaling the transition of the universe into a dark energy-dominated acceleration phase \cite{Visser/2000}. This behavior is consistent with the divergence-free parametric deceleration model and supports the $f(R,T)$ model's alignment with observed cosmic evolution trends.

\section{Conclusion} \label{sec6}

The accelerating expansion of the universe remains a compelling challenge in modern cosmology. Various dynamical dark energy models and modified gravity theories have been proposed to explain this phenomenon, each offering distinct perspectives on cosmic evolution. Despite significant progress, the quest for a comprehensive and consistent description of the accelerating universe continues, motivating further theoretical and observational investigations. In this work, we investigated the divergence-free parametric form of the deceleration parameter within the simplest non-minimal matter-geometry coupling in the $f(R,T)$ gravity framework, where $R$ is the Ricci scalar and $T$ is the trace of the energy-momentum tensor \cite{Harko/2011}. 

Specifically, we considered the simplest linear form $f(R,T) = R + 2\lambda T$, where $\lambda$ is a constant. Using the divergence-free parametric form of the deceleration parameter \cite{Mamon/2016,Hanafy/2019}, we derived the Hubble parameter as a function of redshift $z$ and incorporated this solution into the modified Friedmann equations. By employing observational data from OHD and the Pantheon compilation, we obtained precise constraints on the model parameters, including the Hubble parameter $H_0$, the present deceleration parameter $q_0$, and its evolutionary component $q_1$. The best-fit values confirm a smoothly evolving $q(z)$, ensuring a consistent transition between cosmic deceleration and acceleration. Furthermore, we analyzed the evolution of the energy density $\rho$ and the EoS parameter $\omega$ for various values of the coupling parameter $\lambda$, which governs the interaction between matter and geometry. Our results show that smaller $\lambda$ values enhance cosmic acceleration, with the EoS parameter deviating more significantly from the standard $\Lambda$CDM behavior. The divergence-free nature of the model ensures a continuous evolution of $ \omega$, transitioning from the matter-dominated phase ($\omega \approx 0$) to the present accelerating phase ($\omega < -\frac{1}{3}$). Finally, we examined the energy conditions to assess the physical viability of the model. The NEC and DEC remain satisfied throughout cosmic evolution, indicating a physically consistent framework. The SEC is violated at late times, supporting the observed acceleration of the universe \cite{Visser/2000}. The influence of $\lambda$ is evident in moderating the effective energy density and pressure, thereby shaping the universe's dynamical evolution. Furthermore, we have focused on positive values of $\lambda$ in our analysis, as these correspond to scenarios where the matter-geometry coupling enhances cosmic acceleration, offering a viable alternative to DE models. Negative values of $\lambda$ lead to negative energy density, which is unphysical within the framework of our model.

In comparison to other cosmological models, our findings provide a distinct perspective on cosmic acceleration. While the $\Lambda$CDM model assumes a constant vacuum energy, our model introduces a dynamical contribution from the matter-geometry interaction, leading to deviations from standard cosmic expansion. Compared to other modified gravity approaches, such as $f(R)$ or $f(Q)$ gravity \cite{MK1, MK2, MK3, MK4, MK5, MK6, MK7, MK8}, our model offers a minimal yet effective extension to GR that maintains physical viability and remains consistent with observational data. The divergence-free deceleration parameter ensures a smooth cosmic evolution across all redshifts, including $z \to -1$, avoiding potential singularities or becoming invalid at certain redshifts, as seen in alternative parameterizations \cite{Riess/2004,Santos/2011,Nair/2012,Xu/2008,Gong/2006,Campo/2012,Mamon/2018,Mamon/2017}. The coupling parameter $\lambda$ plays a crucial role in controlling deviations from $\Lambda$CDM, making this approach an interesting alternative for understanding the interaction between matter and modified gravity effects. Future studies could extend this analysis to include more precise observational datasets and explore potential implications for structure formation and cosmic anisotropies.

\section*{Acknowledgments}
The authors sincerely appreciate the editor and the anonymous referee for their valuable comments and constructive suggestions, which have significantly improved the clarity and quality of this manuscript. Their interesting feedback has helped refine our analysis and presentation.

\section*{Data Availability Statement}
This article does not introduce any new data.

\end{document}